\documentclass[12pt,preprint]{aastex}

\shorttitle{Ammonia in the Egg nebula}
\shortauthors{Chiu et al.}

\begin{document}
\title{Shock-enhanced ammonia emission in the Egg nebula}

\author{ 
Dinh-V-Trung\footnote{on leave from Center for Quantum Electronics, Institute of Physics,
Vietnamese Academy of Science and Technology, 10 DaoTan, ThuLe, BaDinh, Hanoi, Vietnam}, 
P.J. Chiu\footnote{Institute of Astronomy, National Central University
No. 300, Jhongda Rd, Jhongli City, Taoyuan County 320, Taiwan},
Jeremy Lim}
\affil{Institute of Astronomy and Astrophysics, Academia Sinica\\
P.O Box, 23-141, Taipei 106, Taiwan\\}
\email{trung@asiaa.sinica.edu.tw, pjchiu@asiaa.sinica.edu.tw, jlim@asiaa.sinica.edu.tw}

\begin{abstract}
We present high angular resolution observations of the NH$_3$(1,1), (2,2) and (3,3) inversion 
transitions from the Egg Nebula, the archetypical proto-planetary nebula. 
The spatial distribution and kinematics of the emission in all three
lines show four distinct components or lobes that are aligned with the polar
and equatorial directions. The kinematics of the NH$_3$ emission is also found to follow
a clear pattern: redshifted emission in the South and West and blueshifted emission in the
North and East. The morphology and spatial kinematics of NH$_3$ emission are shown to have 
strong similarity to that observed previously in molecular hydrogen emission
and CO emission which arise from the shocked molecular gas. 
We also find that the higher lying inversion transition NH$_3$ (2,2) and (3,3) are stronger in the
polar direction in comparison to the lower transition NH$_3$ (1,1).
We conclude that the NH$_3$ emission traces the warm molecular gas, which is shocked and heated
by the interaction between the high velocity outflows and the
surrounding envelope. 
The presence of strong ammonia emission associated with the shock fronts and the lack of the emission
at the center of the nebula indicate that the abundance of ammonia is significantly enhanced by 
shocks, a situation very similar to that found in outflows from protostars.
\end{abstract}


\section{Introduction}

The Egg nebula (CRL 2688) is widely considered as the prototype of proto-planetary nebulae (PPN),
a class of stars in the rapid transition phase between the AGB
and planetary nebulae. During the PPN phase the central post-AGB star contracts and gradually becomes
hotter, but is not yet hot enough to ionize its surrounding envelope. Using
high resolution spectroscopic observations of the scattered stellar light
Klochkova et al. (2000) determine a spectral type F5Ia
for the central star with an effective temperature of T$_{\rm eff}$=6500K. 
The abundance analysis reveals the enrichment of C and N together with
strong enhancement of slow neutron capture (s-process) elements, as expected for the
carbon rich post-AGB star at the center of the Egg nebula. 
Jura \& Kroto (1990) suggest that the central star left the AGB a few hundreds years ago. The last episode of
high mass loss rate at the end of the AGB phase, commonly referred as the superwind
phase, has created a massive circumstellar envelope around the central post-AGB star. As a result, the Egg nebula
is a strong source of continuum and molecular emissions. 
Numerous and strong molecular lines
have been detected by Young et al. (1992), Truong-Bach et al. (1988).
From continuum emission Jura et al. (2001)
inferred a gas mass of $\sim$0.2 M$_\odot$ scaled to the distance of 420 pc recently determined by 
Ueta et al. (2006). 

High spatial resolution optical images of CRL 2688 reveal 
a pair of bipolar lobes seen as search-light beams emanating
from the central (obscured) star and oriented perpendicular to a prominent dark lane (Sahai et al. 1998a). 
This conspicuous dark lane has been commonly
interpreted as an equatorial disk of cold dust. High angular resolution images of hot 
molecular hydrogen gas emission at 2.2 $\mu$m reveal shocked gas, 
tracing the strong interaction between the fast outflows and
the slow wind (Sahai et al. 1998b). Surprisingly, hot molecular hydrogen emission is seen in both bipolar lobes
as well as in the equatorial plane far beyond the dark lane, indicating the presence of 
multiple fast collimated outflows. The presence of high velocity outflows has also been inferred
from the detection of high velocity wings in the strong emission lines of CO and CS molecules by
Young et al. (1992).

Indeed, high spatial resolution mapping of CO J=2--1 emission by Cox et al. (2000) reveals the 
presence of several pairs of collimated fast molecular outflows along both the bipolar axis and
the equatorial plane. Surprisingly, these molecular outflows can be traced back to a common origin, presumably the
location of the central post-AGB star in the Egg nebula. 
Cox et al. (2000) suggest that
the observed CO emission does not constitute the actual fast collimated outflows, but rather
molecular gas swept up by even faster collimated outflows that comprise mainly 
atomic or ionized gas. In contrast to its appearance in the optical, no equatorial disk is the
Egg nebula is seen in the CO J=2--1 by Cox et al. (2000). 
High spatial resolution observation of the high density tracer HCN J=1--0 
by Bieging \& Nguyen-Q-Rieu (1996) indicates higher gas density along both 
the bipolar lobes and the equatorial plane, consistent with the hypothesis that the molecular gas
gas is swept up by the interaction with the fast outflows. 

Ammonia (NH$_3$) is known to exist in many different environments. The inversion transitions of NH$_3$ which
are closely spaced in frequency can be detected and imaged at high angular resolution almost
simultaneously, allowing quite straightforward excitation analysis (Ho \& Townes 1983). 
The detected inversion lines from metastable levels J=K can be used as a natural thermometer 
to constrain the temperature of the
molecular gas (Walmsley \& Ungerechts 1983).
The (1,1) and (2,2) inversion
transitions in the Egg nebula were first detected in single dish observations of Nguyen-Q-Rieu et al. (1984). 
Follow-up observations by Truong-Bach et al. (1988) also detected the higher excitation lines (3,3) and (4,4). 
Subsequent high angular resolution observation of the NH$_3$(1,1) line by
Nguyen-Q-Rieu (1986) indicates that NH$_3$ is distributed in a disk-like structure
lying along the dark lane with a faint spur extending from the center of
the disk to the north. Nguyen-Q-Rieu et al. (1986) suggest that the
NH$_3$ emission is confined in a disk or more probably in a torus
oriented perpendicular to the optical lobes. The torus, which has a diameter of $\sim$12\arcsec\,
and a thickness less than 3\arcsec, is expanding at a velocity of
$\sim$15 kms$^{-1}$. The low sensitivity and low velocity resolution of the observation by
Nguyen-Q-Rieu et al. (1986),
however, preclude any firm conclusion on the spatial kinematics of the 
ammonia in the Egg nebula.

In this paper we report high angular resolution observations of NH$_3$ inversion lines
(1,1), (2,2) and (3,3) to determine the spatial kinematics and to probe the physical
conditions inside the molecular envelope of the Egg nebula.

\section{Observations and Data Reduction}

We carried out the observations of the Egg nebula in the NH$_3$(1,1),
NH$_3$(2,2) and NH$_3$(3,3) inversion lines
using the Very Large Array (VLA\footnote{The National Radio Astronomy Observatory is a facility of the National 
Science Foundation operated under a cooperative agreement by Associated Universities, Inc.}). The NH$_3$(3,3) line was
observed on March 6, 2003 and NH$_3$(1,1) and NH$_3$(2,2) were
observed simultaneously on July 18, 2004. All the observations
were done in the most compact configuration of VLA. 
The phase center of our observations
was $\alpha_{j2000}$=21h02m18.6s, $\delta_{j2000}$=36d41m37.8s, which is the same as the
peak of the continuum emission observed previously by Cox et al. (2000). To observe
these lines we configured the VLA correlator in the 1A mode with
12.5 MHz bandwidth in 64 channels, thus providing a  channel width of 195.3 kHz 
or a velocity resolution of 2.45 kms$^{-1}$ over a useful velocity coverage of $\sim$150 kms$^{-1}$. 
We monitored the nearby quasar 21095+35330 at frequent intervals (every 5 minutes or so) to correct for
the antennas gain variations caused primarily by atmospheric fluctuations. The stronger quasars 2253+161 was 
used to correct for the shape of the bandpass and its variation with time. The absolute flux 
scale of our observations was determined from observation of standard quasar 0137+331 (3C48). 
The data were edited, calibrated and mapped using AIPS data reduction
package. The calibrated visibilities are then Fourier transformed to form the DIRTY images. 
The natural weighting is used throughout.  The DIRTY images were deconvolved
using normal clean algorithm implemented in AIPS. In order to get better
signal-to-noise ratio we smooth the data to a velocity resolution of 4.9 kms$^{-1}$ per
channel. The rms noise level are $\sim$2.0, 1.5 and 1.5 mJy
beam$^{-1}$ for the NH$_3$(1,1), (2,2) and (3,3) line, respectively. The
difference of rms level in three lines is because of the
integration time and weather conditions are different. 
Table 1 provides a summary of our VLA observations.
\begin{table}
\caption{Summary of the VLA observations}
\begin{tabular}{llll}\hline
            &  NH$_3$(1,1)   &  NH$_3$(2,2)   & NH$_3$(3,3)    \\ \hline
Obs. date        &  July 18, 2004 &  July 18, 2004 & March 06, 2003 \\
Int. time &  1 hour & 1 hour & 1 hour  \\
Clean beam    &  3\arcsec.5 x 3\arcsec.1 & 4\arcsec.1x3\arcsec.4 & 3\arcsec.6x3\arcsec \\
                    &  PA = 44$^\circ$.3       & PA = 74$^\circ$.7 & PA = 44$^\circ$.5 \\
$\Delta$V (kms$^{-1}$)          &  4.9  & 4.9 & 4.9 \\
noise level     & 2.0   & 1.5 & 1.5\\ 
(mJy beam$^{-1}$) & & & \\
\hline
\end{tabular}
\end{table}
\section{Results}
Using the VLA we successful imaged the NH$_3$(1,1), (2,2), and (3,3)
inversion transitions in the Egg nebula. In Figure 1, 2 and 3 we show the channel maps of 
the NH$_3$ emission. The emission in each of the
three transitions cover a velocity range between --74 to --4 kms$^{-1}$. The morphology
of the emission is broadly consistent with the lines forming in an expanding environment because
the emission is most compact at the extreme blueshifted and redshifted velocity channels and
becomes progressively more extended in velocity channels around the systemic velocity of 
V$_{\rm LSR}$=--35 kms$^{-1}$ inferred by Truong-Bach et al. (1988) from the observations of a number of
molecular lines in the Egg nebula
. 
We integrate all the emission above 2$\sigma$ level in the channel maps of the (1,1), (2,2) and (3,3) lines.
The resulting total intensity profiles are presented in Figure 4. Clearly the
strength and the shape of all three NH$_3$ lines are almost identical. The line shape is not
the usual parabola, which is typical for an optically thick line formed in an expanding 
spherical circumstellar envelope. Instead, the line profiles have a triangular shape, reflecting the
more peculiar kinematics and spatial distribution of ammonia in the Egg nebula. These three lines of ammonia were
observed previously using the single dish telescope by Nguyen-Q-Rieu et al. (1984) and 
Truong-Bach et al. (1988). Using a conversion factor of $\sim$0.73 Jy K$^{-1}$ appropriate 
for the Effelsberg 100m radio telescope at the frequencies of the NH$_3$ lines, we estimate that
the peak flux in the (1,1) and (3,3) lines is $\sim$60 mJy, similar to that seen in our 
VLA observations (see Figure 4). The strength of the (2,2) line as observed by Nguyen-Q-Rieu et al. (1984) is slightly
weaker than the other two lines. In our VLA observations we find that the strength of the (2,2) line is
the almost the same as that for the (1,1) and (3,3) lines. It is likely that the relative weakness of the (2,2)
in the single dish observation might be due to either calibration and/or the low S/N of the single dish data. 
Therefore, we conclude that our VLA observations have recovered all the flux in 
all three inversion lines of ammonia.

It's well known that each of the inversion lines of ammonia has complex hyperfine structure. However,
the hyperfine splitting is only significant for the (1,1) line (Kukolich 1967).
The hyperfine structure of the (1,1) line consists of the main line group and four groups of satellite
lines, which are shifted in velocity in either direction by $\sim$8 and $\sim$20 kms$^{-1}$. 
Assuming the the hyperfine lines are optically thin and form under LTE condition,
the intensity of the satellite groups at $\pm$20 kms$^{-1}$ is about 20\% of that
for the main line (Kukolich 1967, Rydbeck et al. 1977). Because the peak flux of the (1,1) line is about
$\sim$10$\sigma$ in the velocity channels --43.6 kms$^{-1}$ and --48.6 kms$^{-1}$, 
where the rms noise level $\sigma$ = 2 mJy beam$^{-1}$, the contribution to the intensity of
the line in other velocity 
channels such as --23.9 kms$^{-1}$ and --28.8 kms$^{-1}$ due to the hyperfine splitting is approximately is 
$\sim$2$\sigma$. This contribution due to hyperfine
structure is modest in comparison to the observed (1,1) lines intensity in the abovementioned velocity channels. 
As a result, we conclude that hyperfine
structure in the (1,1) line does not affect significantly the discussion on the
spatial distrubtion of the emission and also the velocity structure
of the NH$_3$ in the Egg nebula.

Close inspection of the channel maps reveals that the spatial distribution and 
kinematics of the emission in three transitions of NH$_3$ are very similar. 
We can clearly separate the ammonia emission into four components having 
distinct spatial-kinematics, which will be referred 
as the north, south, east and west lobes. The
north and south lobes are most easily seen in the blueshifted and redshifted velocity
channels between --53 to --68 kms$^{-1}$ and --4 to --24 kms$^{-1}$, respectively. These
components extend to the north and south of the center of the nebula along the
bipolar axis as defined in the optical images. 
In the channel maps of all three lines, only the north lobe can be seen as distinct.
The south lobe is a protrusion connected to the stronger east lobe.
The east and west lobes can be seen prominently in the velocity channels between --29 to --48 
kms$^{-1}$ straddling the systemic
velocity. These components are stronger in intensity and aligned in the equatorial 
direction of the nebula. In all three lines the intensity of the emission from the
east lobe is dominant over other components. We note that the east and west lobes are located on either
side of the center of the nebula where there is a general lack of emission. 
That can be seen notably in the central velocity channel at --34 kms$^{-1}$
of NH$_3$ (1,1) and (3,3) lines shown in Figures 1 and 3.

We integrate all the emission stronger than 2$\sigma$ in the channel maps to produce the
integrated intensity maps of all three NH$_3$ lines. The maps are shown in Figures 5, 6 and 7. 
The NH$_3$ emission in all three lines shows a cross-like shape, especially in NH$_3$(3,3).
The cross-like shape is the natural result of the presence of four components identified in
the channel maps. The total spatial extent of the emission is about $\sim$14\arcsec\, in both the polar
direction and the equatorial direction. We note that the north and south
lobes appear distinct and more prominent in ortho-NH$_3$(3,3) transition than in lower lying para-NH$_3$ 
transitions (1,1) and (2,2). Comparing to previous observation of Nguyen-Q-Rieu et al. (1986), we
can see that the dramatic improvement in sensitivity of the VLA allows us to detect the fainter emission in
the north and south lobes and provides the full picture of the NH$_3$ distribution in the Egg nebula. 

To investigate the kinematics of the NH$_3$ emission we form the intensity-weighted 
mean velocity maps of all three lines using all the
emission stronger than 4$\sigma$ level in the channel maps. The resulting maps are also shown
in Figures 5, 6 and 7. In all three lines the the north and south
lobes exhibit opposite mean velocity, i.e. the emission in the north lobe is entirely blue-shifted and 
that in the south lobe is entirely red-shifted. These two lobes in the polar direction are
almost mirror symmetric in spatial morphology and mean velocity with respect to the center
of the nebula. In the equatorial direction, the kinematics of the east and west lobes is 
more complicated. That is most clearly seen in the mean velocity maps of NH$_3$(1,1) and (3,3) lines.
Most of the emission in the east lobe is blueshifted with respect to the systemic velocity. 
The northern portion of the east lobe is, however, moving at slower velocities almost the same as
the systemic velocity. The kinematics of the west lobe is almost mirror symmetric with respect to the
of the east lobe, i.e most of the emission is redshifted with respect to the systemic
velocity with the southern portion moving at an average velocity close to the systemic
velocity. Most likely the east and west lobes consist of multiple kinematic sub-components with
very different spatial kinematics. This point will be discussed further in the next section.
\section{Discussion}
\subsection{Line ratios}
The inversion transitions of NH$_3$ have been widely used to gauge the physical conditions in
different environments, especially in starforming regions. Therefore our obserations of 
multilines in the Egg nebula can be used  to infer the physical conditions of the
molecular gas through the excitation analysis. More specifically, the gas temperature can be
estimated from the line ratios together with some plausible assumptions. We form the line ratio maps (2,2)/(1,1)
and (3,3)/(1,1) by convolving the channel maps to the same angular resolution of 4\arcsec.1
and then producing the integrated intensity maps and then the line ratio maps.
The line ratio maps are shown in Figures 8 and 9. Surprisingly, the spatial distribution
of the two line ratios is very similar. Although the east and west lobe are stronger in
the integrated intensity maps, the line ratios there are generally low, ranging from about 0.3 to 0.9
for both NH$_3$(2,2)/(1,1) and NH$_3$(3,3)/(1,1). In constrast, the line ratios are generally higher in the
polar direction, ranging from 0.8 up to 1.4 for both cases. Because the excitation of higher lying 
transitions NH$_3$(2,2) and (3,3) requires hot and dense gas, high line ratios
in the polar direction suggests that the molecular gas in the polar lobes is both hotter and denser in comparison to
the molecular gas present in the equatorial plane.

If we assume that the NH$_3$ inversion lines are optically thin, the opacity ratio for any pair of lines 
can be calculated as follows (Ho \& Townes 1983): 
\begin{eqnarray}\label{rotation temperature}
    \frac{\tau(J',K')}{\tau(J,K)} = \frac{\nu^2(J',K')}{\nu^2(J,K)}
    \frac{\Delta\nu(J',K')}{\Delta\nu(J,K)} \frac{T_{ex}(J,K)}
    {T_{ex}(J',K')} \frac{\mid\mu(J',K')\mid^2}{\mid\mu(J,K)\mid^2}
    \frac{g(J',K')}{g(J,K)} \nonumber \\
     \times exp\left\{\frac{-\Delta E(J',K'; J,K)}{kT_R(J',K'; J,K)}
    \right\},
\end{eqnarray}
where T$_{\rm ex}$(J,K) is the excitation temperature of the inversion line, 
$\mid\mu(J,K)\mid^2$ = $\mu^2 K^2$/[J(J+1)] are the dipole
matrix elements, $g(J,K)$ is the statistical weight, $g(J,K)$ = $g_{op}(2J + 1)$, with
$g_{op}$ = 1 for para-ammonia levels
(J = 1, 2, 4, 5, 7, 8,...) and $g_{op}$ = 2 for ortho-ammonia levels (J = 0, 3,
6, ...). $\Delta E(J',K';J,K)/k$ is the energy difference between two
states (J',K') and (J,K). In the optically thin limit, we can use
the line ratio to estimate the rotation temperature T$_R$ between states (J,K) and (J',K'), which is a good
indicator of the gas temperature T$_{\rm K}$. The rotation temperature
can be calculated from the following expression (Menten \& Alcolea 1995): 
\begin{equation}\label{ErotT31}
T_{R}(J',K';J,K)=\left[\frac{E(J',K')-E(J,K)}{k}\right]
\left\{{\rm ln}\left[ \frac{\int T_b(J,K)dv}{\int T_{b}(J',K')dv} 
\frac{J(J+1)K'^2(2J+1)}{J'(J'+1)K^2(2J+1)} \frac{g'_{op}}{g_{op}}\right]\right\}^{-1}
\end{equation}
The energy difference between (2,2) and (1,1) states is 41.2 K and that for (3,3) and (1,1) states
is 100.3 K, respectively (Menten \& Alcolea 1995).
From the values of the (2,2)/(1,1) ratio mentioned above we estimate that the rotation temperature T$_{\rm R}$(2,2;1,1)
ranges between $\sim$21 to 46 K in the east and west lobes. In the polar direction the rotation temperature
T$_{\rm R}$(2,2;1,1) can reach as high as 90 K because of the higher line ratio. Similarly, we find
that the rotation temperature T$_{\rm R}$(3,3;1,1) is in the range between 32 to 50 K in the east and west lobes
and reaches higher value up to 62 K in the polar direction. We note that the high rotation temperatures derived here are 
broadly consistent with our optically thin assumption for the inversion
lines of NH$_3$. As we can see from the channel maps of (1,1), (2,2) and (3,3) lines, the brightness temperature
is in the range 1 to 4 K, eventhough the emitting regions are spatially resolved. Thus the inversion
lines are likely optically thin if the excitation temperatures T$_{\rm ex}$(J,K) is comparable to the derived 
rotation temperatures.
We caution, however, that the rotation temperature
between ortho-ammonia state (3,3) and the para-ammonia state (1,1) is of limited meaning because of the 
selection rule that forbids any radiative or collisional transition connecting between the ortho and para states
(Ho \& Townes 1983). 
\subsection{Association of NH$_3$ emission with shocked molecular gas}
The complex morphology and spatial kinematics of ammonia emission indicate
that ammonia emission does not originate from the unshocked remnant AGB wind. Instead,
there are several lines of evidences suggesting that the emission is associated with the shocked molecular
gas produced by the interaction between high velocity jets and the slowing expanding
shell ejected when the central star was still on the AGB phase.

From high angular resolution observations of CO J=2--1 line, Cox et al. (2000) identify a series
of high velocity molecular outflows in the Egg nebula. These outflows are oriented in the polar axis and in
the equatorial direction. More interestingly, these outflows can be traced back to a common origin close
to the center of the nebula. In Figures 4, 5 and 6 we overlay the location of the molecular outflows
identified by Cox et al. (2000) on the integrated intensity maps of NH$_3$ emission. These
outflows form a cross-like shape, very similar to the morphology of the NH$_3$ emission,
and spatially coincide with the location of the four lobes traced by the
NH$_3$ emission. We also find that the NH$_3$ emission covers a similar velocity range (between --74 to --4 kms$^{-1}$) 
as the CO J=2--1 line and exhibits similar spatial kinematics. Namely, the CO emission in the polar direction appears 
blueshifted in the north and redshifted in the south, as expected from the mean velocity maps of NH$_3$. Similarly,
The CO J=2--1 emission associated with outflows in the equatorial direction exhibits blueshifted emission in the east
and redshifted in the west, exactly the same trend seen in the intensity weighted mean velocity maps of NH$_3$ emission
shown in Figures 5, 6 and 7. 

The kinematics of strong molecular hydrogen emission at 2.2 $\mu$m, which is commonly believed to trace the shocked gas, 
has been shown by Kastner et al. (2001) to be similar to that seen in CO J=2--1 emission. In addition, 
in the equatorial direction, the prominent and extended molecular hydroge emission in the east, 
which exhibits blueshifted velocities, 
can be separated into three distinct sub-components with different 
spatial kinematics. Among the three sub-component E2, E3 and E4 in the terminology of Sahai et al. (1998b), the 
E4 sub-component located to the north is moving more slowly than other sub-components. Comparison with the mean velocity
maps of NH$_3$ emission we also see that the northern portion of the east lobe has lower average velocity, almost the same
as the systemic velocity. Kastner et al. (2001) propose that the spatial kinematics of the molecular hydrogen 
emission could result from rotation in the equatorial plane of the Egg nebula. A similar suggestion was 
made early by Bieging \& Nguyen-Q-Rieu (1996) based on their observations of the HCN J=1--0 line. However, 
our current observations, which have similar spatial and velocity resolution, do not add any new 
information to either support or disprove the suggestion of rotation in the equatorial plane in the Egg nebula.

The similarity in spatial distribution and kinematics observed in molecular hydrogen emission and NH$_3$ 
inversion transitions strongly suggest that the NH$_3$ emission originates from the shocked molecular gas.
We note, however, that the molecular hydrogen emission arises from molecular gas at temperature at least as high
as 1000 -- 2000 K, whereas the NH$_3$ emission is coming from much lower temperature gas as suggested by the
low rotation temperatures of the inversion lines estimated in the previous section. This would imply a large
temperature gradient within the shocked region and the emission lines from either molecular hydrogen or NH$_3$
would trace the temperature range most favorable to the their excitation.  
\subsection{Is the abundance of ammonia enhanced in shocks ?}
The strong ammonia emission associated with the shocked molecular gas and the lack of emission
at the center of the Egg nebula suggest that most of the ammonia did not formed initially
in the AGB wind but is produced in the shocked gas. Generally, NH$_3$ is not thought to form
through gas-phase chemistry in the dusty wind of AGB stars. In chemical models for 
circumstellar envelopes, NH$_3$ is either not considered (Glassgold et al. 1986, Cherchneff et al. 1993)
or assumed to formed initially in the stellar atmosphere and injected into the wind (Millar \& Herbst 1994).
Willacy \& Cherchneff (1998) considered the effect of shock processing on the molecular abundances for
the case of the prototypical carbon rich AGB star IRC+10216. However, the number abundance of NH$_3$ (relative to H$_2$) 
even in that case remains very low, up to only about a few 10$^{-11}$.
Nejad et al. (1990) and Flower et al. (1995) suggest that the NH$_3$ molecules might form through surface chemistry
on dust grains and remain on the grain surface until the release by the passage of shocks. In their model, a shock 
velocity in the range 10 to 15 km s$^{-1}$ would heat up the gas and lead to the release of NH$_3$ from 
the dust grain mantles. The theoretical calculations of Burton et al. (1992) indicate that a minimum shock velocity
of about 10 kms$^{-1}$ is neccessary to excite the molecular hydrogen 2.2 $\mu$m line. The observations of strong molecular
hydrogren 2.2 $\mu$m emission in the Egg nebula (Sahai et al. 1998b) suggest that the strength of the shocks in the 
Egg nebula is comparable or even stronger than needed in the abovementioned chemical models in order to produce 
the abundance enhancement of NH$_3$.
In addition, the relatively high rotation temperatures estimated from the ratio of ammonia inversion lines also
favors this scenario in the Egg nebula. We note that similar observations of the enhanced ammonia emission in the bipolar outflows
from young stellar objects. For L1155 outflow, Tafalla \& Bachiller (1995) found very good spatial coincidence between ammonia emission
and the distribution of SiO J=2--1 line, which is considered as a good tracer of the shocked gas. Follow up observations
by Umemoto et al. (1999) revealed the emission from highly excited (5,5) and (6,6) inversion lines. The high rotational
temperature of ammonia molecules as well as the significant departure of the ortho--para ratio from the statistical
equilibrium value point to the evaporation of NH$_3$ from the surface of dust grains following the passage of shocks. 

Given the relatively well constrained geometry and
physical conditions inside the Egg nebula, more elaborate models should be attempted to 
explore further the formation and excitation of ammonia in the shock regions.
\section{Conclusion}
In this paper we have imaged at high angular resolution the emission of NH$_3$(1,1), (2,2) and (3,3) inversion 
lines from the Egg Nebula. We find that
the spatial distribution and kinematics of the emission in all three
lines show four distinct components or lobes that are aligned with the polar
and equatorial directions. The kinematics of the NH$_3$ emission is also found to follow
a clear pattern: redshifted emission in the South and West and blueshifted emission in the
North and East. The morphology and spatial kinematics of NH$_3$ emission are shown to have 
strong similarity to that observed previously in molecular hydrogen emission
and CO emission which arise from the shocked molecular gas. 
We also find that the higher lying inversion transition NH$_3$ (2,2) and (3,3) are stronger in the
polar direction in comparison to the lower transition NH$_3$ (1,1).
We conclude that the NH$_3$ emission traces the warm molecular gas, which is shocked and heated
by the interaction between the high velocity outflows and the
surrounding envelope. 
From the association of NH$_3$ emission with the shocked molecular gas and the lack of the emission
at the center of the nebula we conclude that the abundance of ammonia is significantly enhanced by 
shocks, a situation very similar to that found in outflows from protostars.
\acknowledgments
We thank an anonymous referee for insightful and constructive comments that help to improve the
presentation of our paper.
We also thank the VLA staff for their help with the observations. 
This research has made use of NASA's Astrophysics Data System Bibliographic Services
and the SIMBAD database, operated at CDS, Strasbourg, France.

\newpage

\begin{figure*}[ht]
\plotone{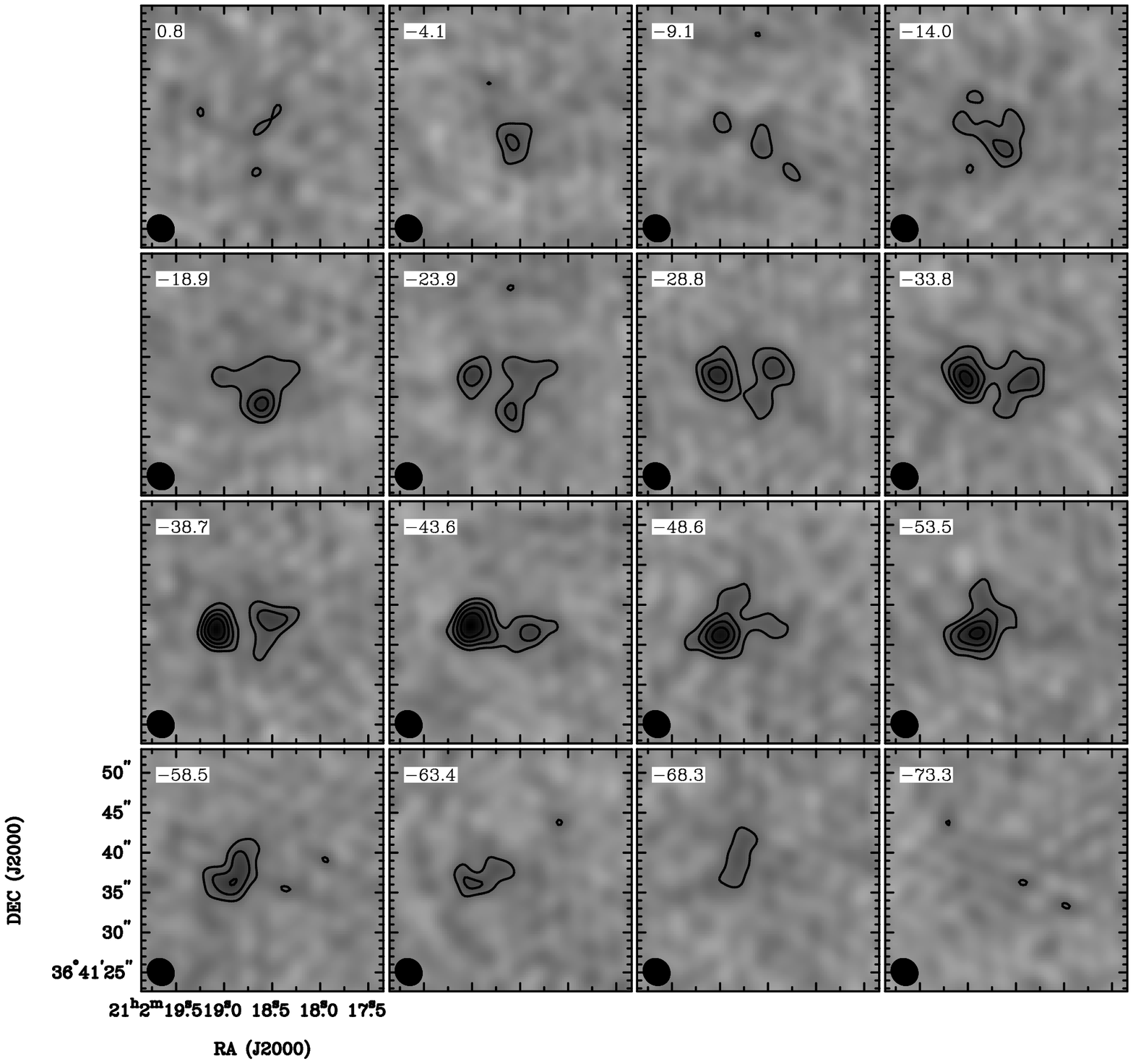}
\caption{Channel maps of the NH$_3$(1,1) emission from the Egg nebula in contour and grayscale. The contour levels
are (3, 5, 7, 9, 12)$\sigma$ where the rms noise level $\sigma$ = 2.0 mJy beam$^{-1}$.
The synthesize beam of 3\arcsec.5x3\arcsec.1 is shown in the lower left of the upper left frame. The conversion factor
between the brightness temperature and the flux of the NH$_3$(1,1) emission is 5 mJy K$^{-1}$.}
\label{fig1}
\end{figure*}

\begin{figure*}[ht]
\plotone{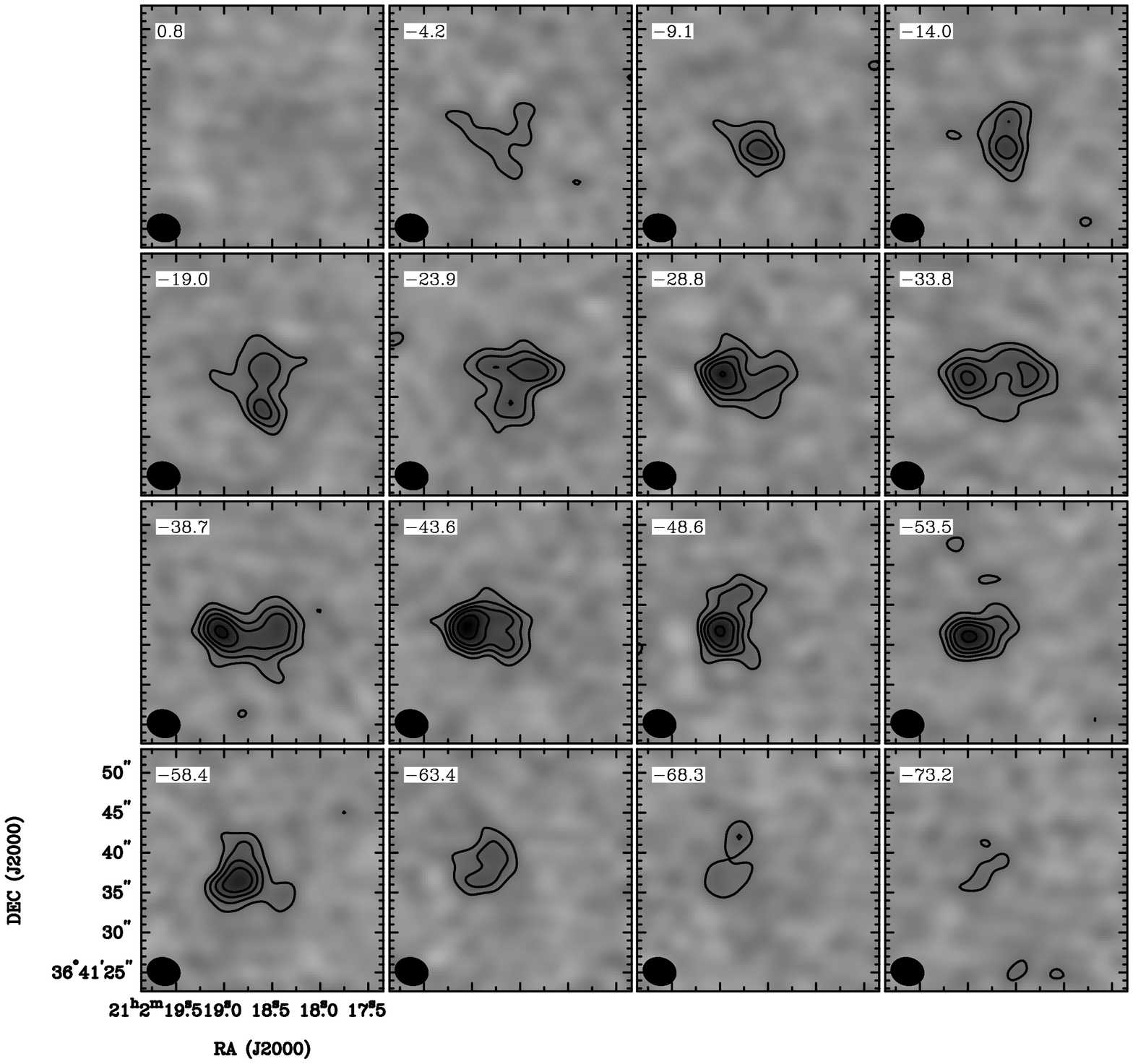}
\caption{Channel maps of the NH$_3$(2,2) emission from the Egg nebula in contour and grayscale. The contour levels
are (3, 5, 7, 9, 12)$\sigma$ where the rms noise level $\sigma$ = 1.5 mJy beam$^{-1}$.
The synthesize beam of 4\arcsec.1x3\arcsec.4 is shown in the lower left of the upper left frame. The conversion factor
between the brightness temperature and the flux of the NH$_3$(2,2) emission is 6.4 mJy K$^{-1}$.}
\label{fig2}
\end{figure*}

\begin{figure*}[ht]
\plotone{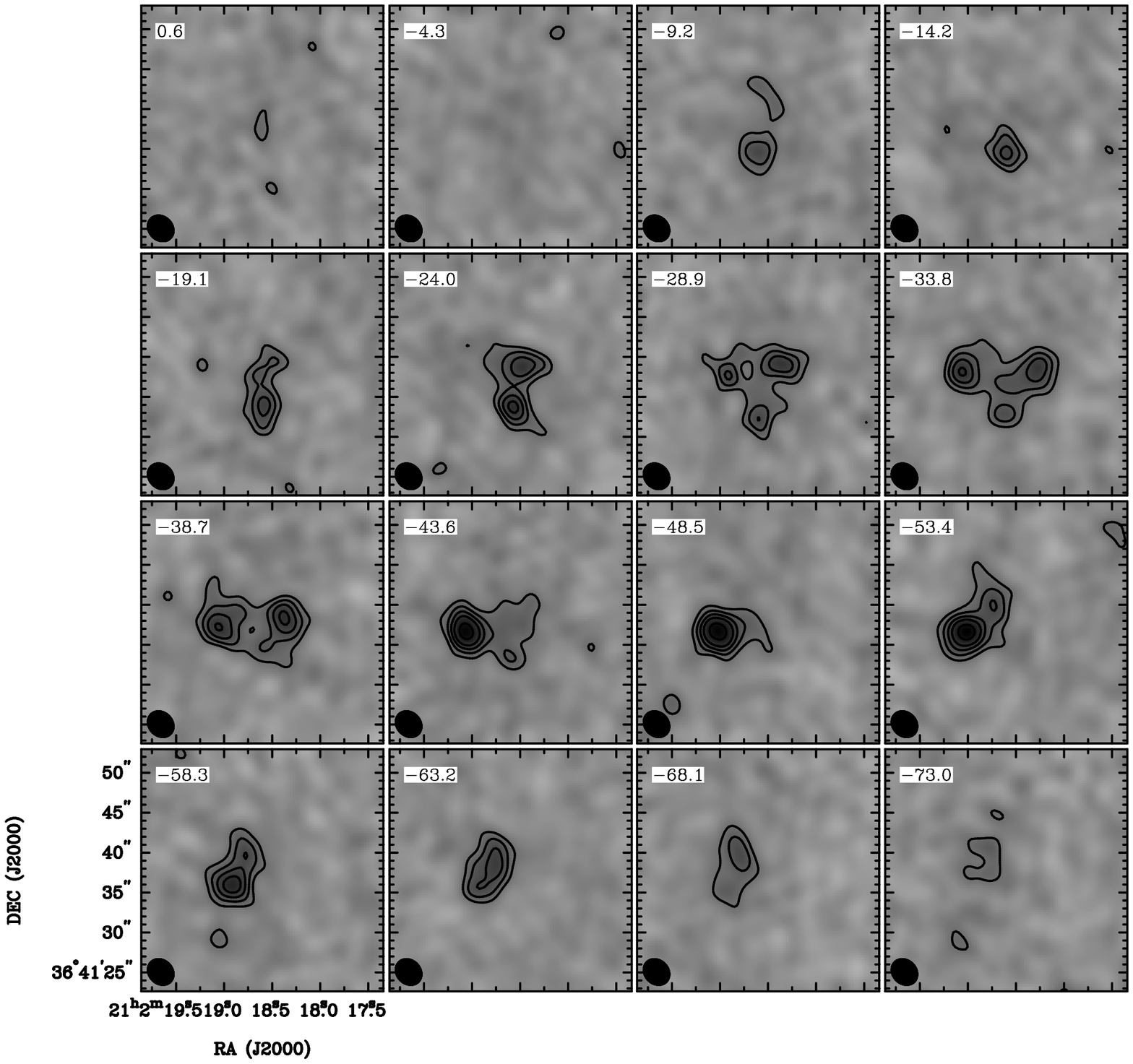}
\caption{Channel maps of the NH$_3$(3,3) emission from the Egg nebula in contour and grayscale. The contour levels
are (3, 5, 7, 9, 12)$\sigma$ where the rms noise level $\sigma$ = 1.5 mJy beam$^{-1}$.
The synthesize beam of 3\arcsec.6x3\arcsec is shown in the lower left of the upper left frame. The conversion factor
between the brightness temperature and the flux of the NH$_3$(3,3) emission is 5 mJy K$^{-1}$.}
\label{fig3}
\end{figure*}

\begin{figure*}[ht]
\plotone{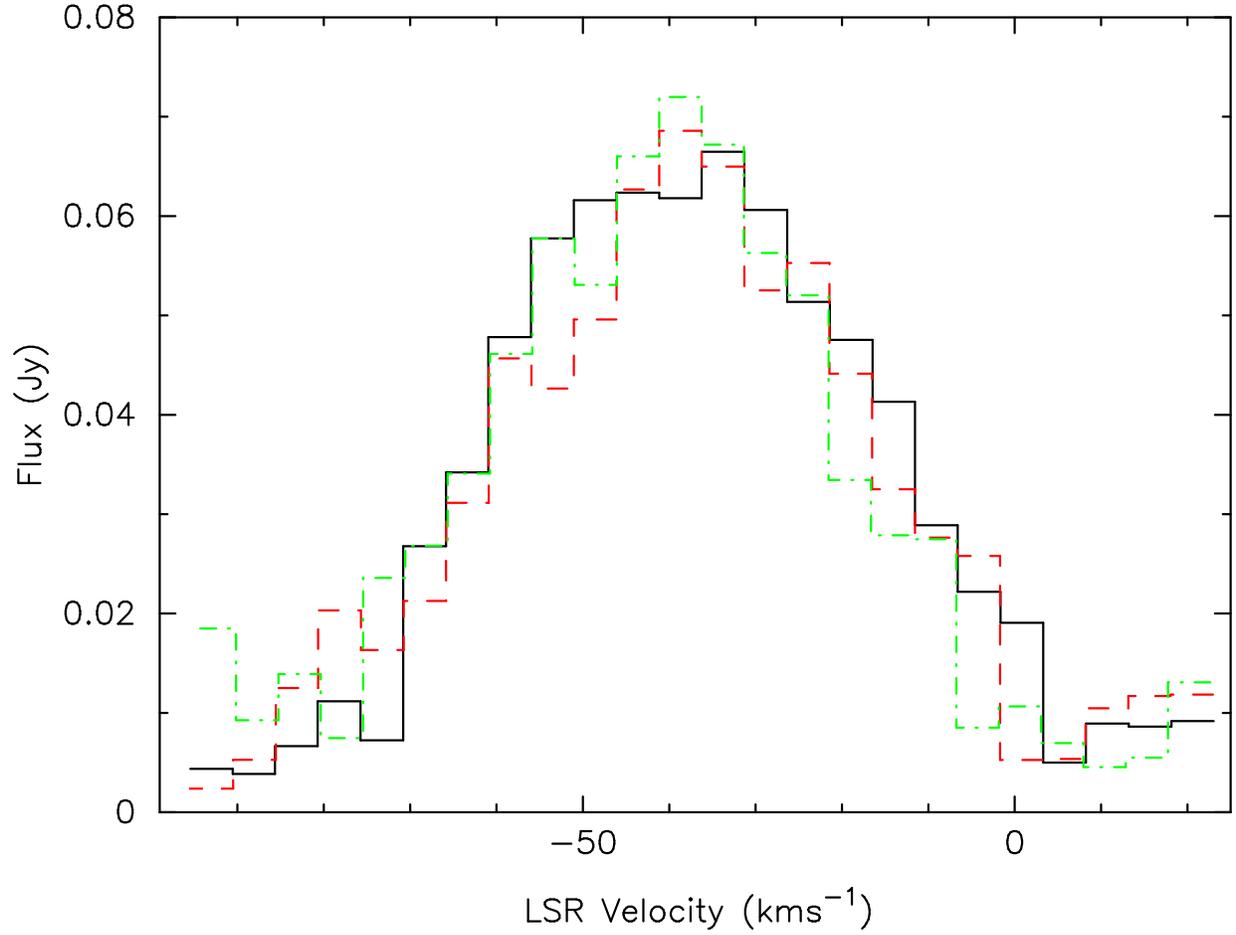}
\caption{Total intensity profiles of the NH$_3$ (1,1) transition, shown in black solid line, (2,2) transition,
shown in dashed line, and (3,3) transition, shown in dashed-dotted line, from the Egg nebula.}
\label{fig4}
\end{figure*}

\begin{figure*}[ht]
\plotone{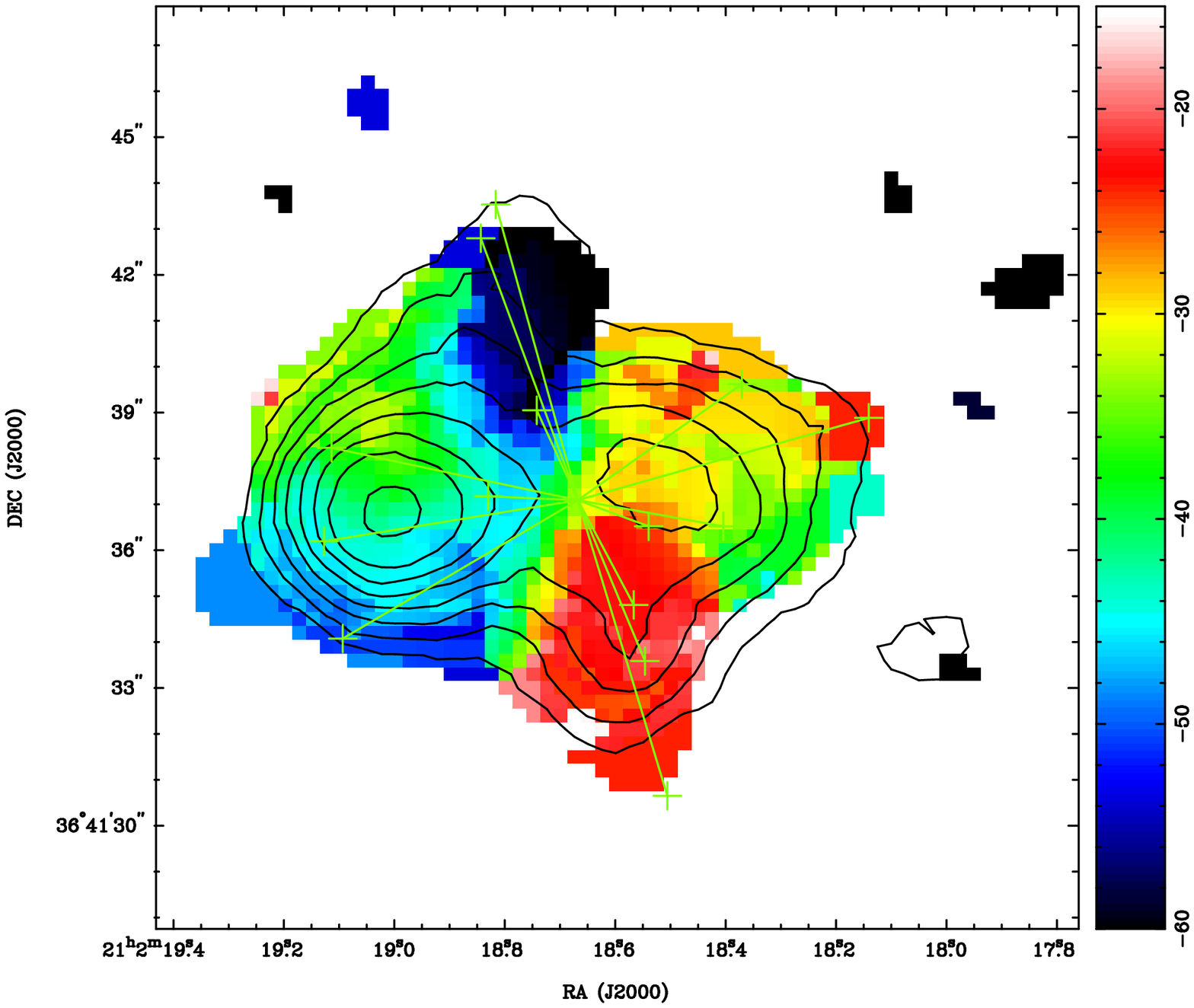}
\caption{The integrated intensity map of the NH$_3$(1,1) line from the Egg nebula. The contour
levels start from (0.2 to 0.8 in step of 0.2 and 0.95)x0.9 Jy kms$^{-1}$. The high velocity jets identified
by Cox et al. (2000) are marked with solid lines.}
\label{fig5}
\end{figure*}

\begin{figure*}[ht]
\plotone{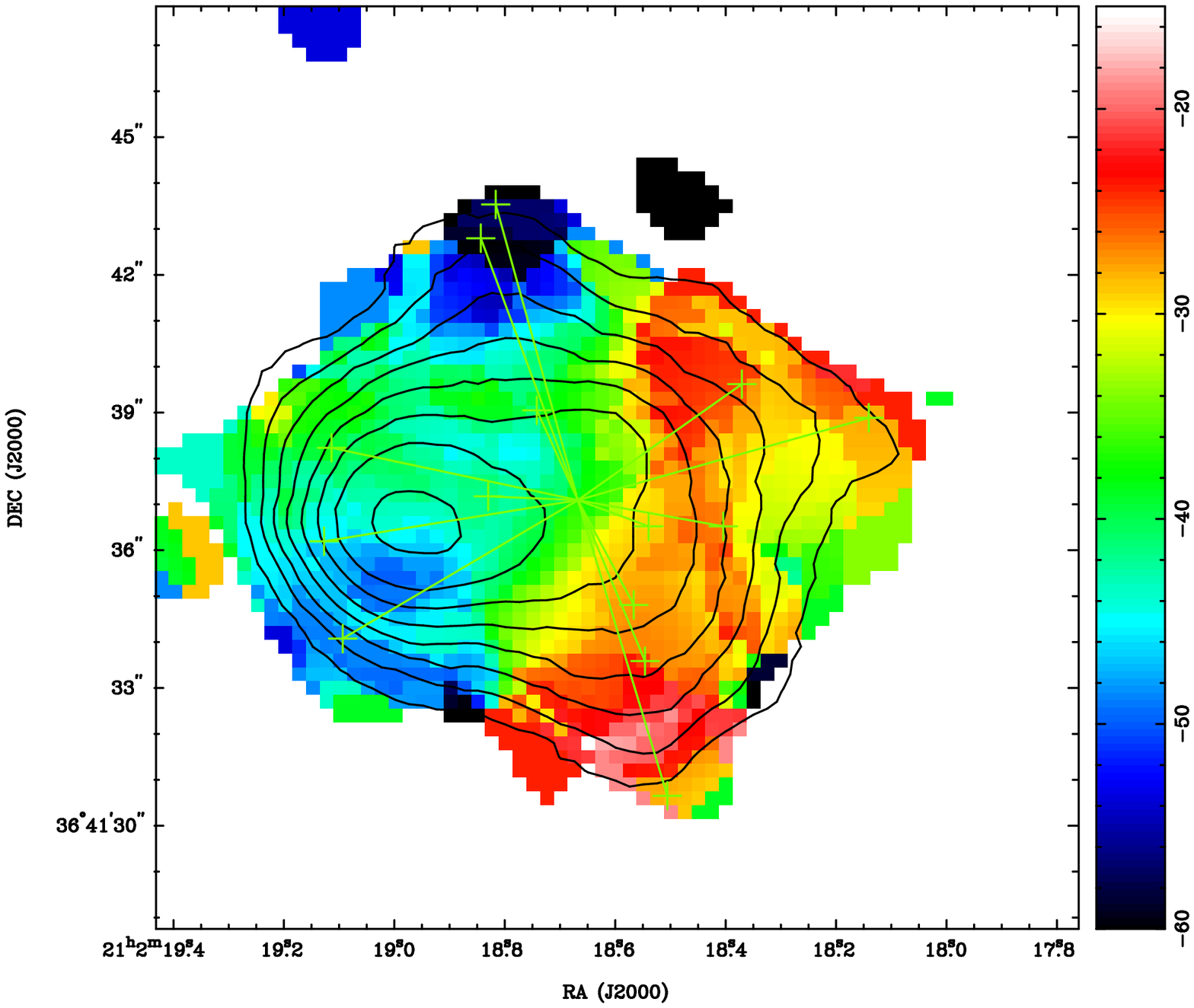}
\caption{The integrated intensity map of the NH$_3$(2,2) line from the Egg nebula. The contour
levels start from (0.1 to 0.8 in step of 0.2 and 0.95)x0.83 Jy kms$^{-1}$. The high velocity jets identified
by Cox et al. (2000) are marked with solid lines.}
\label{fig6}
\end{figure*}

\begin{figure*}[ht]
\plotone{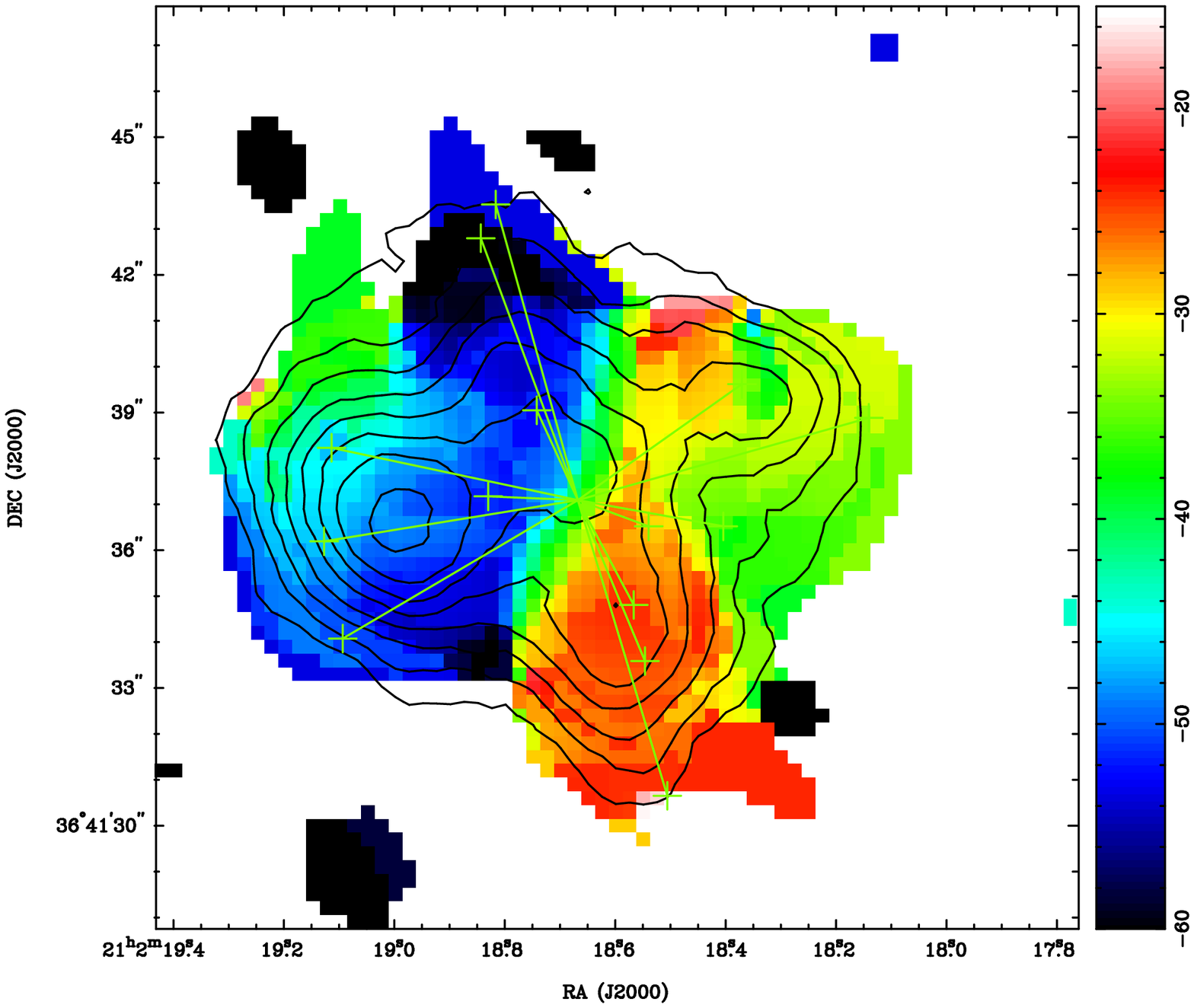}
\caption{The integrated intensity map of the NH$_3$(3,3) line from the Egg nebula. The contour
levels start from (0.2 to 0.8 by step of 0.2 and 0.95)x0.67 Jy kms$^{-1}$. The high velocity jets identified
by Cox et al. (2000) are marked with solid lines.}
\label{fig7}
\end{figure*}

\begin{figure*}[ht]
\plotone{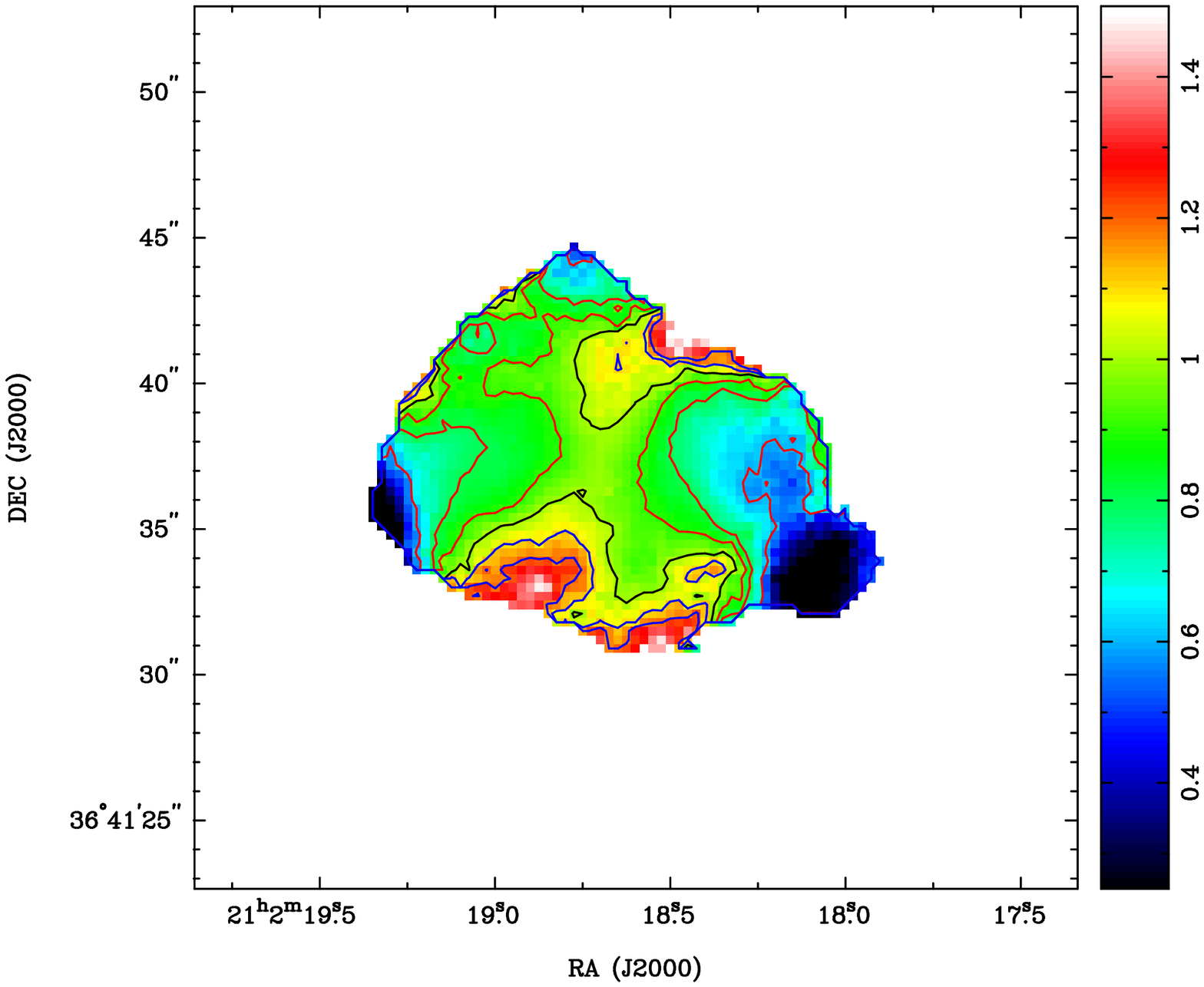}
\caption{The intensity ratio map of the NH$_3$(2,2)/(1,1) line from the Egg nebula. The contour
levels are 0.6, 0.8, 0.9, 1, 1.1 and 1.2. The map has an angular resolution
of 4\arcsec.1.}
\label{fig8}
\end{figure*}

\begin{figure*}[ht]
\plotone{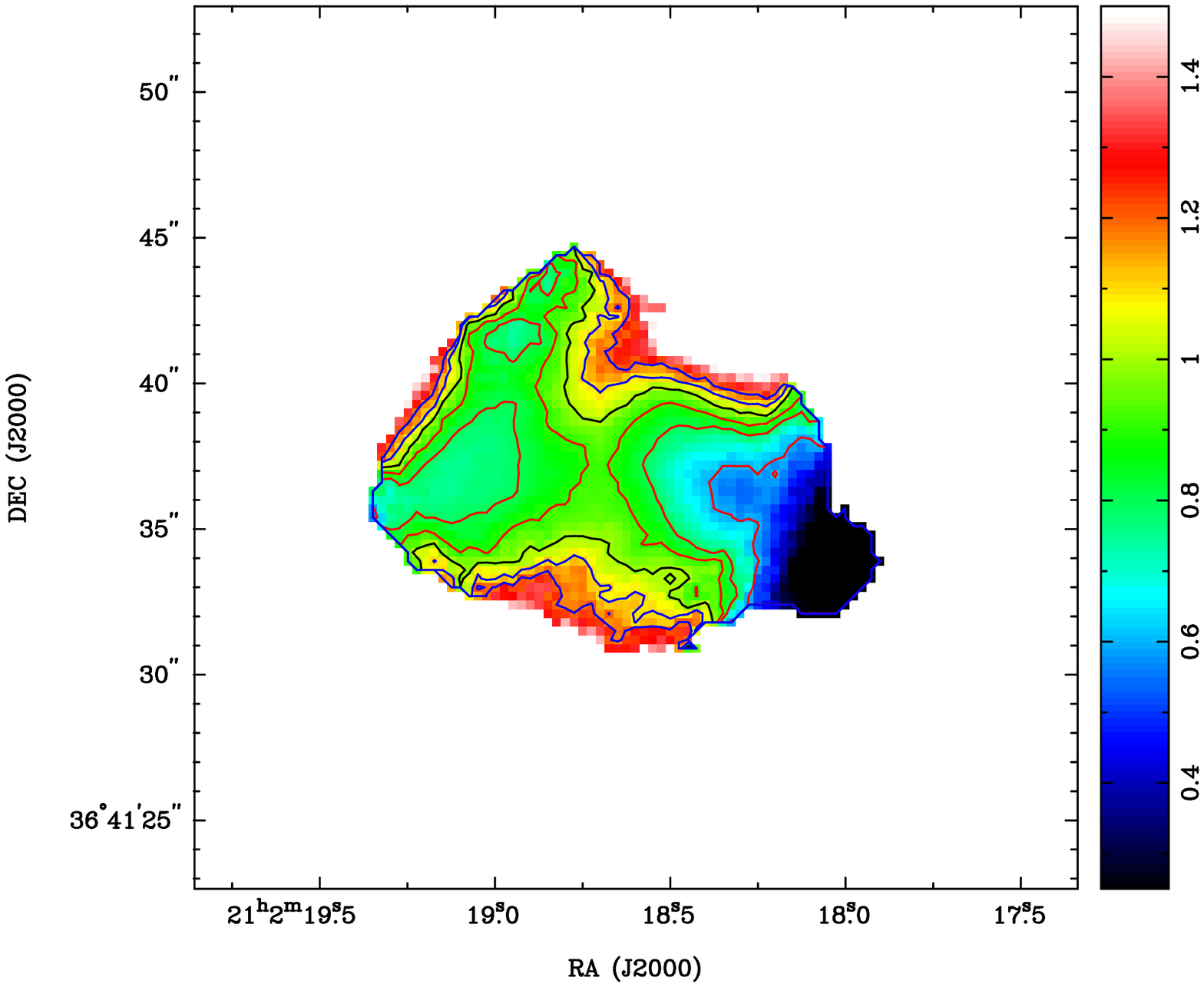}
\caption{The intensity ratio map of the NH$_3$(3,3)/(1,1) line from the Egg nebula. The contour
levels are 0.6, 0.8, 0.9, 1, 1.1 and 1.2. The map has an angular resolution 
of 4\arcsec.1.}
\label{fig9}
\end{figure*}

\end{document}